\documentclass[prb,aps,showpacs,onecolumn]{revtex4}
\usepackage{amsfonts}
\usepackage{graphicx}
\usepackage{amsmath}
\usepackage{bm}

\input{tcilatex}

\begin{document}

\title{Bohlin-Arnold-Vassiliev's duality and conserved quantities}
\author{Y. Grandati, A. B\'erard and H. Mohrbach}
\affiliation{Laboratoire de Physique Mol\'eculaire et des Collisions, ICPMB, IF CNRS
2843, Universit\'e Paul Verlaine, Institut de Physique, Bd Arago, 57078
Metz, Cedex 3, France}

\begin{abstract}
Bohlin-Arnold-Vassiliev's duality transformation establishes a
correspondence between motions in different central potentials. It offers a
very direct way to construct the dynamical conserved quantities associated
to the isotropic harmonic oscillator (Fradkin-Jauch-Hill tensor) and to the
Kepler's problem (Laplace-Runge-Lenz vector).
\end{abstract}

\maketitle


\section{Introduction}

The study of the relation between the classical motions of the harmonic
oscillator and the Kepler problem has a long history initiated in Hooke and
Newton's works\cite{grant}. In 1911, Bohlin showed that this relation can be
formulated in terms of conformal mapping\cite{bohlin}. This conformal
transform is at the core of the Levi-Civita's regularization scheme\cite%
{levi} of which tri-dimensional generalization has been achieved by
Kustaanheimo and Stiefel\cite{KS}. In fact, as noted by Needham\cite%
{needham1,needham2} two years before Bohlin's paper, Kasner\cite{kasner} had
established a more general duality law relating pairs of power law
potentials (but the result has only been published in 1913). This relation
has been rediscovered and generalized by Arnold and Vassiliev\cite%
{arnvas1,arnvas2} in 1989 and quite simultaneously by Hojman and al.\cite%
{hojman}. In fact, the generalization of Kasner's result had already been
obtained by Collas\cite{collas} and implicitly enters into the frame of the
coupling constant metamorphosis of Hietarinta and al.\cite{hietarinta,
rosquist,tsiganov}. Even if we limit ourselves to the classical aspects (The
study of this correspondence in the quantum mechanical frame has an
interesting parallel history that we won't deal with here), numerous
articles have been published on this subject during the last fifteen years 
\cite{grant,mittag,wu,mitchell1,stump1,stump2,hall}. In all what follows, we
stay in a purely Newtonian frame. On the basis of simple but general
arguments concerning the complex representation of 2D-motions, we show that
the Bohlin-Arnold-Vassiliev's dual correspondence appears very naturally as
the relevant transformation of motions among all combinations of analytic
change of coordinates and Euler-Sundman reparameterization. We consider the
effect of this duality on conserved quantities. In the specific cases of
Hooke's and Kepler's problems, we recover in a very simple and direct way
the correspondence established by Nersessian and al.\cite{ners1,ners2,ners3}
between the associated additional dynamical conserved quantities. This
permits to point out the essentially obvious character of Fradkin-Jauch-Hill
tensor's and Laplace-Runge-Lenz vector's conservation.

\section{Motions transformations}

\subsection{Complex formulation}

Let's consider a planar motion $\overrightarrow{r}\left( t\right) =\binom{%
x(t)}{y(t)}_{(O,\overrightarrow{u_{x}},\overrightarrow{u_{y}})}$ for a
particle of mass $m$ submitted to a potential $U(\overrightarrow{r})$,
eventually singular at the origin. We will adopt a complex formulation and
represent the position by its corresponding affix $z(t)=x(t)+iy(t)$, the
potential being viewed as a real valued function, $U(z,\overline{z})$. The
gradient of such a real valued function of $z$ and $\overline{z}$, is given
by: 
\begin{equation}
\quad \overrightarrow{\nabla }U(\overrightarrow{r})=2\frac{\partial U(z,%
\overline{z})}{\partial \overline{z}}
\end{equation}

where $\frac{\partial }{\partial \overline{z}}=\frac{1}{2}\left( \frac{%
\partial }{\partial x}+i\frac{\partial }{\partial y}\right) =\overline{\frac{%
\partial }{\partial z}}$. In particular, for a central potential: $U(z,%
\overline{z})=U\left( \left| z\right| \right) =U\left( r\right) $ and $\frac{%
\partial U(r)}{\partial z}=\frac{dU(r)}{dr}\frac{z}{r}$.

The equation of motion takes the form:

\begin{equation}
\overset{..}{z}+\frac{2}{m}\frac{\partial U(z,\overline{z})}{\partial 
\overline{z}}=0  \label{eqmouv1}
\end{equation}

where the dot represents the time $t$ derivative.

Multiplying both sides by $\overset{.}{\overline{z}}$ and taking the real
part of the resulting quantity, we obtain a total derivative:

\begin{equation}
\frac{\overset{.}{\overline{z}}\overset{..}{z}+\overset{.}{z}\overset{..}{%
\overline{z}}}{2}++\frac{1}{m}\left( \overset{.}{\overline{z}}\frac{\partial
U(z,\overline{z})}{\partial \overline{z}}+\overset{.}{z}\frac{\partial U(z,%
\overline{z})}{\partial z}\right) =0
\end{equation}

Integrating this identity we recover the conservation of energy:

\begin{equation}
E=\frac{1}{2}m\left| \overset{.}{z}\right| ^{2}+U(z,\overline{z})=\frac{1}{2}%
m\left\| \overset{.}{\overrightarrow{r}}(t)\right\| ^{2}+U(\overrightarrow{r}%
)  \label{energie1}
\end{equation}
If $\overrightarrow{A}$ and $\overrightarrow{B}$ are two vectors in the $(O,%
\overrightarrow{u_{x}},\overrightarrow{u_{y}})$ plane, the complex affix of $%
\overrightarrow{A}\times \overrightarrow{B}$ is given by the real quantity $%
\func{Im}\left( \overline{A}B\right) $, $A$ and $B$ being the complex
affixes of $\overrightarrow{A}$ and $\overrightarrow{B}$ respectively (Note
that this real quantity doesn't correspond to a vector in the $(O,%
\overrightarrow{u_{x}},\overrightarrow{u_{y}})$ plane but which is
orthogonal to this last).

The angular momentum $\overrightarrow{L}\left( t\right) =m\overrightarrow{r}%
\left( t\right) \times \overset{.}{\overrightarrow{r}}\left( t\right)
=L\left( t\right) \overrightarrow{u_{z}}$ admits then as complex
correspondent $L\left( t\right) =m\func{Im}\left( \overline{z}(t)\overset{.}{%
z}(t)\right) =\frac{m}{2i}\left( \overline{z}\overset{.}{z}-\overset{.}{%
\overline{z}}z\right) $.

Moreover:

\begin{equation}
\overset{.}{L}\left( t\right) =\frac{m}{2i}\left( \overline{z}\overset{..}{z}%
-\overset{..}{\overline{z}}z\right) =i\left( \overline{z}\frac{\partial U(z,%
\overline{z})}{\partial \overline{z}}-\frac{\partial U(z,\overline{z})}{%
\partial z}z\right)
\end{equation}

which implies that for a central potential $\overset{.}{L}\left( t\right) =0$%
.

\subsection{Conformal change of coordinates}

Consider first an arbitrary conformal change of coordinates:

\begin{equation}
\ z=f(w),\ \overline{z}=\overline{f(w)}=\overline{f}(\overline{w})
\end{equation}

Then : 
\begin{equation}
\frac{\partial }{\partial \overline{z}}=\left( \frac{\partial \overline{z}}{%
\partial \overline{w}}\right) ^{-1}\frac{\partial }{\partial \overline{w}}=%
\frac{1}{\overline{f}^{\left( 1\right) }(\overline{w})}\frac{\partial }{%
\partial \overline{w}}=\frac{1}{\overline{f^{\left( 1\right) }(w)}}\frac{%
\partial }{\partial \overline{w}}
\end{equation}

where $f^{\left( n\right) }(z)=\frac{\partial ^{n}f(z)}{\partial z^{n}}$.

After substitution, Eq.(\ref{eqmouv1}) takes the following form:

\begin{equation}
\overset{..}{\ w}+\left( \overset{.}{\ w}\right) ^{2}\log \left( f^{\left(
1\right) }(w)\right) ^{\left( 1\right) }+\frac{2}{m}\frac{1}{\left|
f^{\left( 1\right) }(w)\right| ^{2}}\frac{\partial \widetilde{U}(w,\overline{%
w})}{\partial \overline{w}}=0  \label{eqmouv2}
\end{equation}

where $\widetilde{U}(w,\overline{w})=U\left( f(w),\overline{f}(\overline{w}%
)\right) $.

\subsection{Euler-Sundman's reparameterization}

The previous equation~(\ref{eqmouv2}) incorporates now a term of first order
in time derivative, which has an artificial character. In order to suppress
it, we can envisage to change the parameterization for the motion $w(t)$. A
global change of parameterization will be inefficient. Rather, we choose a
change of parameterization having a local structure, that is an
Euler-Sundman's reparameterization of the type:

\begin{equation}
dt=\left| g\left( w\right) \right| ^{2}ds  \label{eulertransf}
\end{equation}
where $g\left(w\right) $ is an, a priori, arbitrary analytic function. This
form ensures in particular that the correspondence between the initial and
the new (fictitious) time is one-to-one and increasing.

Starting from Eq.(\ref{eqmouv2}), it is straightforward to show that the
equation for the motion $w(s)$\ is now given by:

\begin{equation}
w^{\prime \prime }+\left( w^{\prime }\right) ^{2}\left( \log \frac{f^{\left(
1\right) }(w)}{g\left( w\right) }\right) ^{\left( 1\right) }-\overline{%
\left( \log g\left( w\right) \right) ^{\left( 1\right) }}\left| w^{\prime
}\right| ^{2}+\frac{2}{m}\frac{\left| g\left( w\right) \right| ^{4}}{\left|
f^{\left( 1\right) }(w)\right| ^{2}}\frac{\partial \widetilde{U}(w,\overline{%
w})}{\partial \overline{w}}=0  \label{eqmouv3}
\end{equation}

A priori, the structure of this last equation is still more complicated than
the previous one, since it incorporates two kinds of terms depending on
first derivative in time: the first one depending on $\left( w^{\prime
}\right) ^{2}$ as in Eq.(\ref{eqmouv2})) and the second one, depending on $%
\left| w^{\prime }\right| ^{2}$. But the gain becomes readily clear. Indeed,
the freedom in the choice of $g\left( w\right) $ allows to expect to
suppress the first one. As for the problem induced by the second, as we just
go to see, it is solved by the use of energy conservation identity.

\subsection{Energy conservation}

Expressed in terms of $w$ and $s$, the energy conservation equation~(\ref%
{energie1}) gives:

\begin{equation}
\left| w^{\prime }\right| ^{2}=\frac{2}{m}\left( E-\widetilde{U}(w,\overline{%
w})\right) \frac{\left| g\left( w\right) \right| ^{4}}{\left| f^{\left(
1\right) }(w)\right| ^{2}}  \label{energie2}
\end{equation}

We are then in position to replace in Eq.(\ref{eqmouv3}) the contribution
containing $\left| w^{\prime }\right| ^{2}$ by another one, depending only
of the position. We obtain:

\begin{equation}
w^{\prime \prime }+\left( w^{\prime }\right) ^{2}\left( \log \left( \frac{%
f^{\left( 1\right) }(w)}{g\left( w\right) }\right) \right) ^{\left( 1\right)
}+\frac{2}{m}\frac{\left| g\left( w\right) \right| ^{4}}{\left| f^{\left(
1\right) }(w)\right| ^{2}}\left( \frac{\partial \widetilde{U}(w,\overline{w})%
}{\partial \overline{w}}-\overline{\left( \log g\left( w\right) \right)
^{\left( 1\right) }}\left( E-\widetilde{U}(w,\overline{w})\right) \right) =0
\end{equation}

\subsection{ Reparameterization's choice}

In order to eliminate the contribution containing $\left( w^{\prime }\right)
^{2}$, we have to fix $g\left( w\right) $ such that $g\left( w\right)
=Cf^{\left( 1\right) }\left( w\right) ,\ C\in \mathbb{C}$. The simplest
choice, which we'll adopt in the sequel, is to take $C=1$ and: 
\begin{equation}
g\left( w\right) =f^{\left( 1\right) }\left( w\right)  \label{reparam}
\end{equation}

Then Eq.(\ref{eqmouv4}) becomes simply :

\begin{equation}
w^{\prime \prime }+\frac{2}{m}f^{\left( 1\right) }(w)\frac{\partial }{%
\partial \overline{w}}\left( \overline{f^{\left( 1\right) }(w)}\left( 
\widetilde{U}(w,\overline{w})-E\right) \right) =0  \label{eqmouv5}
\end{equation}

Defining :

\begin{equation}
V_{f}(w,\overline{w})=\left| f^{\left( 1\right) }(w)\right| ^{2}\left( 
\widetilde{U}(w,\overline{w})-E\right) +V_{0}  \label{potdual1}
\end{equation}

where $V_{0}$ is an arbitrary real constant, we finally obtain :

\begin{equation}
w^{\prime \prime }+\frac{2}{m}\frac{\partial V_{f}(w,\overline{w})}{\partial 
\overline{w}}=0  \label{eqmouv6}
\end{equation}

In other words, $w\left( s\right)$ is the motion of a particle of mass $m$
in the potential $V_{f}(w,\overline{w})$ (if we choose as supplementary
additive term an arbitrary holomorphic function $V_{0}(w)$, we are led to
the same equation but the corresponding potential $V_{f}(w,\overline{w})$ is
no more real valued).

As for the energy of this dual system, it's given by :

\begin{equation}
\widetilde{E}=\frac{m}{2}\left| w^{\prime }\right| ^{2}+V_{f}(w,\overline{w}%
)=V_{0}
\end{equation}

Relation~(\ref{potdual1}) becomes finally:

\begin{equation}
\left| f^{\left( 1\right) }(w)\right| ^{2}=\frac{\widetilde{E}-V_{f}(w,%
\overline{w})}{E-\widetilde{U}(w,\overline{w})}=\frac{\widetilde{E}-V_{f}(w,%
\overline{w})}{E-U(f\left( w\right) ,\overline{f}\left( \overline{w}\right) )%
}  \label{reparam2}
\end{equation}

We then obtain a functional equation to which is submitted the
transformation $f$ linking the motions in the potentials $U$ and $V$
respectively.

In conclusion, among all the transformations of a 2D-motion in a potential $%
U $ combining a conformal (i.e. analytical) change of coordinates $z=f(w)$
and an Euler-Sundman reparameterization $dt=\left| g\left( w\right) \right|
^{2}ds$, those for which the transformed motion is an autonomous 2D-motion
submitted to a potential $V$, necessarily satisfy Eq.(\ref{reparam}) and Eq.(%
\ref{reparam2}).

\section{Bohlin-Arnold-Vassiliev's duality}

\subsection{Arnold-Vassiliev's potentials}

We consider the case where the initial potential $U(\overrightarrow{r})$
takes the form $U(\overrightarrow{r})=U(z,\overline{z})=A\overline{u(z)}%
u(z)=A\left| u(z)\right| ^{2}$ , where $u(z)$ is an analytical function on $%
\mathbb{C}^{\ast }$ and $A\in \mathbb{R}$. We'll call such a potential, an
Arnold-Vassiliev's potential\cite{arnvas2}. It is easy to show that the only
central Arnold-Vassiliev's potentials are the power law potentials,
corresponding to $u(z)\sim z^{\frac{\nu }{2}},\ \nu \in \mathbb{R}$, that is:

\begin{equation}
U(z,\overline{z})=A\left| z\right| ^{\nu},\quad \nu ,A\in \mathbb{R}
\label{potcentral2}
\end{equation}

If $\widetilde{u}(w)=\left( u\circ f\right) (w)$ , we have, by the
transformation associated to $f$, an associated dual potential of the form
(see Eq.(\ref{reparam2})):

\begin{equation}
\widetilde{E}-V(w,\overline{w})=\left| f^{\left( 1\right) }(w)\right|
^{2}\left( E-A\left| \widetilde{u}(w)\right| ^{2}\right)  \label{reparam3}
\end{equation}

Let's ask for which type of conformal transformation $f$, the corresponding
image potential $V$ is also an Arnold-Vassiliev's potential $V(w,\overline{w}%
)=B\left| v(w)\right| ^{2}$.

Inserting in Eq.(\ref{reparam3}) and deriving successively with respect to $%
w $ and $\overline{w}$, we obtain:

\begin{equation}
\frac{AE}{B\widetilde{E}}\frac{\left| v(w)\right| ^{2}}{\left| \widetilde{u}%
(w)\right| ^{2}}=\left( \frac{\widetilde{E}-B\left| v(w)\right| ^{2}}{%
E-A\left| \widetilde{u}(w)\right| ^{2}}\right) ^{2}=\left| f^{\left(
1\right) }(w)\right| ^{4}  \label{reparam4}
\end{equation}

If we choose $B$ and $\widetilde{E}$ such that $\frac{AE}{B\widetilde{E}}=1$%
, this gives the following simple constraint :

\begin{equation}
\frac{\left| v(w)\right| }{\left| \widetilde{u}(w)\right| }=\left| f^{\left(
1\right) }(w)\right| ^{2}
\end{equation}

which is clearly satisfied when:

\begin{equation}
v(w)=f^{\left( 1\right) }(w)=\frac{1}{u(f(w))}=\frac{1}{u(z)}
\label{condarnvas1}
\end{equation}

that is when $w=f^{-1}(z)=\int u(z)dz$.

Then Eq.(\ref{reparam4}) gives:

\begin{equation}
1=\frac{\widetilde{E}-B\left| f^{\left( 1\right) }(w)\right| ^{2}}{E\left|
f^{\left( 1\right) }(w)\right| ^{2}-A}
\end{equation}

which, jointly with $\frac{AE}{B\widetilde{E}}=1$, implies: 
\begin{equation}
\widetilde{E}=-A,B=-E  \label{condarnvas2}
\end{equation}

The transformations satisfying Eq.(\ref{condarnvas1}) and Eq.(\ref%
{condarnvas2}) let stable the set of Arnold-Vassiliev's potentials. Since
they are involutions, they can be seen as duality transformations:

\begin{equation}
\left( E,A\left| u(z)\right| ^{2}\right) \overset{w=\int u(z)dz}{\rightarrow 
}\left( -A,\frac{-E}{\left| u(z)\right| ^{2}}\right) \overset{\zeta =\int
v(w)dw=z}{\rightarrow }\left( E,A\left| u(z)\right| ^{2}\right)
\label{arnvas}
\end{equation}

The Arnold-Vassiliev's potentials appear then as the more natural class of
potentials for which the above combination of an analytical change of
coordinates and an Euler-Sundman reparameterization is a dual correspondence.

\subsection{Power-law potentials}

Let's consider the particular case of power-law potentials: $u(z)=z^{\frac{%
\nu }{2}},\ U(z,\overline{z})=k\left| z\right| ^{\nu }$, at fixed energy $E$%
. Eq.(\ref{eqmouv1}) becomes:

\begin{equation}
\overset{..}{z}+\frac{\nu k}{m}\left| z\right| ^{\nu -2}z=0
\end{equation}

The preceding results (see Eq.(\ref{arnvas})) give:

\begin{equation}
w=f^{-1}(z)=\frac{1}{1+\frac{\nu}{2}}z^{\frac{\nu}{2}+1},\quad z=f(w)=\left(
1+\frac{\nu}{2}\right) ^{\frac{1}{1+\frac{\nu}{2}}}w^{\frac{2}{\nu+2}}
\label{arnvaspowerlaw1}
\end{equation}

By the duality transformation considered above, the corresponding image
potential (see Eq.(\ref{potdual2}), Eq.(\ref{condarnvas2}) and note that $%
A=k $) is then given by:

\begin{equation}
v(w)=\left( 1+\frac{\nu }{2}\right) ^{-\frac{\nu }{\nu +2}}w^{-\frac{\nu }{%
\nu +2}},\quad V(w,\overline{w})=B\left| v(w)\right| ^{2}=\widetilde{k}%
\left| w\right| ^{\mu }  \label{potdual3}
\end{equation}

where:

\begin{equation}
\mu =-\frac{\nu }{1+\frac{\nu }{2}},\ \widetilde{k}=-E\left( 1+\frac{\mu }{2}%
\right) ^{\mu }  \label{arnvaspowerlaw2}
\end{equation}

The energy for the dual motion is (see Eq.(\ref{condarnvas2})) $\widetilde{E}%
=-k$ and the equation of the dual image motion takes the form:

\begin{equation}
w^{\prime \prime }+\frac{\mu \widetilde{k}}{m}\left| w\right| ^{\mu -2}w=0
\label{arnvaspowerlaw3}
\end{equation}

We recover for the dual image motion, a motion of energy $\left( -A\right) $%
, submitted to a power law potential of characteristic exponent $\mu $, such
that:

\begin{equation}
\left( 1+\frac{\nu}{2}\right) \left( 1+\frac{\mu }{2}\right) =1
\label{arnvaspowerlaw4}
\end{equation}

and for which the coupling constant is proportional to the opposite of the
initial energy $E$ (and real if $\nu >-2$).

The relation between the two motions will be called
''Bohlin-Arnold-Vassiliev's duality''.

In the special case $\nu =2$, we obtain $\mu =-1$. In other words, the dual
motion of the plane harmonic oscillator (Hooke potential $\frac{1}{2}kr^{2}$%
, energy $E$) is nothing but the Kepler motion (Newton potential $-\frac{E}{2%
}\frac{1}{\rho }$, energy $-\frac{k}{2}$). We recover the usual Levi-Civita
regularizing transformation with the associated change of variables:

\begin{equation}
z=\sqrt{2w},\ w=\frac{1}{2}z^{2}  \label{levi1}
\end{equation}

and reparameterization:

\begin{equation}
ds=2\rho dt=r^{2}dt  \label{levi2}
\end{equation}

where $r=\left| z\right| $ and $\rho =\left| w\right| $.

\section{Duality and conserved quantities}

For a general 2D-central potential we have two real conserved quantities:
the energy $E$, associated to translational invariance in time, and the
angular momentum $L$, associated to rotational invariance. In the case of
power-law potentials, the effect of the duality transformation on the energy
is given by Eq.(\ref{arnvaspowerlaw2}). Concerning the angular momentum, it
is easy to show that (see Eq.(\ref{reparam})):

\begin{equation*}
L=\frac{m}{2i}\left( \overline{z}\overset{.}{z}-\overset{.}{\overline{z}}%
z\right) =\frac{m}{2i}\left( \frac{w^{\prime }}{\overline{\left( \log
f\left( w\right) \right) ^{\left( 1\right) }}}-\frac{\overline{w}^{\prime }}{%
\left( \log f\left( w\right) \right) ^{\left( 1\right) }}\right)
\end{equation*}

which gives for power law potentials (see Eq.(\ref{arnvaspowerlaw1})):

\begin{equation*}
L=\left( 1+\frac{\nu }{2}\right) \widetilde{L}
\end{equation*}

where $\widetilde{L}$ is the angular momentum of the $w(s)$ motion.

Among all the central potentials, only the Kepler system and the isotropic
harmonic oscillator possess true additional conserved quantities which are
the Laplace-Runge-Lenz vector\cite{leach} and the Fradkin-Jauch-Hill tensor 
\cite{jauch,fradkin} respectively. With all the other central potentials are
associated only piecewise conserved vectors\cite{holas,grandati}.

This peculiarity of Kepler and harmonic potentials is at the origin of the
existence of closed orbits for all the bound states\cite{Bertrand,grandati2}.

In complex representation the existence of a specific conserved quantity for
the harmonic oscillator becomes a matter of course. Indeed, if we consider
the equation for a plane motion in a central potential $U(r)$:

\begin{equation}
\overset{..}{z}+\frac{1}{m}\frac{U^{\left( 1\right) }(r)}{r}z=0
\end{equation}

We see immediately that among all these potentials, the only one for which
the corresponding equation is not $\overline{z}$ dependent is the linear
one, that is the one associated to the Hooke potential (i.e. isotropic
harmonic oscillator):

\begin{equation}
\overset{..}{z}+\frac{k}{m}z=0  \label{eqmouvHO}
\end{equation}

This characteristic feature allows the existence of an immediate integrating
factor $\overset{.}{z}$, which after integration leads to the following
complex conserved quantity:

\begin{equation}
\mathcal{T}=m\frac{\left( \overset{.}{z}\right) ^{2}}{2}+k\frac{z^{2}}{2}%
=cste
\end{equation}

We then have:

\begin{equation}
\mathcal{T}=\left( T_{11}-T_{22}\right) +iT_{12}  \label{FJHtensor1}
\end{equation}

where $T_{ij}$ is the well known Fradkin-Jauch-Hill's tensor\cite%
{jauch,fradkin}:

\begin{equation}
T=\left( 
\begin{array}{cc}
T_{11} & T_{12} \\ 
T_{12} & T_{22}%
\end{array}
\right) =\left( 
\begin{array}{cc}
\frac{m}{2}\left( \overset{.}{x}\right) ^{2}+\frac{k}{2}x^{2} & m\frac{%
\overset{.}{x}\overset{.}{y}}{2}+k\frac{xy}{2} \\ 
m\frac{\overset{.}{x}\overset{.}{y}}{2}+k\frac{xy}{2} & \frac{m}{2}\left( 
\overset{.}{y}\right) ^{2}+\frac{k}{2}y^{2}%
\end{array}
\right)  \label{FJHtensor2}
\end{equation}

As for the energy, it is nothing else but the trace of this tensor:

\begin{equation}
E=T_{11}+T_{22}=\frac{1}{2}m\left| \overset{.}{z}\right| ^{2}+\frac{1}{2}%
k\left| z\right| ^{2}  \label{energie3}
\end{equation}

In complex representation, the existence of Fradkin-Jauch-Hill's tensor
appears therefore as a trivial consequence of the linearity of the equation
of motion.

Let's interest ourselves to the interpretation of $\mathcal{T}$ for the
associated dual Kepler motion. In terms of $w$ variable and $s$ parameter
(see Eq.(\ref{levi1}) and Eq.(\ref{levi2})), $\mathcal{T}$ \ takes the form :

\begin{equation}
\mathcal{T}=\frac{m}{4w}\left( \overset{.}{w}\right) ^{2}+kw=m\overline{w}%
\left( w^{\prime }\right) ^{2}+kw  \label{FJHtensor3}
\end{equation}

Since the energy of the dual motion is given by $\widetilde{E}=-\frac{k}{2}=%
\frac{1}{2}m\left| w^{\prime }\right| ^{2}-\frac{E}{2}\frac{1}{\rho }$, we
can write: 
\begin{equation}
\mathcal{T}=mw^{\prime }\left( \overline{w}w^{\prime }-w\overline{w}^{\prime
}\right) +E\frac{w}{\rho }=2imw^{\prime }\widetilde{L}-2\widetilde{k}\frac{w%
}{\rho }  \label{FJHtensor4}
\end{equation}

where $\widetilde{k}=-\frac{E}{2}$.

If we note $\overrightarrow{\rho }=\xi \overrightarrow{u}_{x}+\eta 
\overrightarrow{u}_{y}$ the vector of affix $w$, $i\widetilde{L}w^{\prime }$
is the affix of $\overrightarrow{\widetilde{L}}\times \overrightarrow{\rho }%
^{\prime }$. We then obtain:

\begin{equation}
\mathcal{T}=2\widetilde{k}\mathcal{A}=-E\mathcal{A}  \label{FJHetLRL}
\end{equation}

where:

\begin{equation}
\mathcal{A}=\frac{m}{\widetilde{k}}iw^{\prime }\widetilde{L}-\frac{w}{\rho }
\label{LRL1}
\end{equation}

We immediately recognize the affix of the well-known Laplace-Runge-Lenz
vector\cite{leach} for the Kepler motion $w$:

\begin{equation}
\overrightarrow{A}=\frac{m}{\widetilde{k}}\overrightarrow{\widetilde{L}}%
\times \overrightarrow{\rho }^{\prime }-\frac{\overrightarrow{\rho }}{\rho }
\label{LRL2}
\end{equation}

By Eq.(\ref{FJHetLRL}) we then have a very direct interpretation of the
Laplace-Runge-Lenz vector as the ratio of two conserved quantities
associated to the dual motion of isotropic harmonic oscillator: 
\begin{equation}
\mathcal{A}=-\frac{\mathcal{T}}{E}
\end{equation}

This ensures immediately the conservative character of $\mathcal{A}$.

The complex Newtonian frame adopted here permits to bring out the obvious
character of the conservation Laplace-Runge-Lenz vector as a consequence,
via the dual transform, of the triviality of the Fradkin-Jauch-Hill tensor's
conservation. It is interesting to note that the elementary approach
developed here can be extended to a much more general set of planar motions
conserving only the angular momentum direction\cite{grandati3}. In this
case, the duality connects different classes of motions. For those
presenting closed orbits which satisfy generalized Gorringe-Leach equations,
there exist additional conserved quantities which are linked by a relation
similar to Eq.(\ref{FJHetLRL}).

\end{document}